\documentclass[]{aastex631}

\usepackage{amsmath}	% Advanced maths commands
\usepackage{comment}
\usepackage[T1]{fontenc}

\def\sigmat{\sigma_{\rm T}}
\def\mel{m_{{\rm e}}}
\def\mp{m_{\rm p}}
\def\nel{n_{{\rm e}^-}}
\def\np{n_{\rm p}}
\def\mel{m_{{\rm e}}}
\def\mp{m_{\rm p}}
\def\xs{x_{\rm s}}
\def\hs{h_{\rm s}}
\def\lps{l_{\rm ps}}
\def\lsk{l_{\rm sk}}
\def\rg{r_{\rm g}}
\def\tg{t_{\rm g}}
\def\lsim{\lower.5ex\hbox{$\; \buildrel < \over \sim \;$}}
\def\gsim{\lower.5ex\hbox{$\; \buildrel > \over \sim \;$}}

\begin{document}

\title[Radiatively driven jets]{Numerical Simulation of Radiatively driven Transonic Relativistic Jets}

\author[0000-0002-9036-681X]{Raj Kishor Joshi}
\affiliation{Aryabhatta Research Institute of Observational Sciences (ARIES) \\
Manora Peak  \\
Nainital 263001, India  }
\affiliation{Department of Astronomy, Astrophysics and Space Engineering\\ 
Indian Institute of Technology Indore\\
Khandwa Road, Simrol 453552, India}

\affiliation{Department of Physics, Deen Dayal Upadhyaya Gorakhpur University \\
Gorakhpur, 273009, India  \\
}

\author[0000-0002-2133-9324]{Indranil Chattopadhyay}
\affiliation{Aryabhatta Research Institute of Observational Sciences (ARIES) \\
Manora Peak  \\
Nainital 263001, India  }

\author[0000-0003-2242-8924]{Antonios Tsokaros}
\affiliation{Department of Physics, University of Illinois at Urbana-Champaign\\
Urbana, Illinois 61801, USA}
\affiliation{National Center for Supercomputing Applications, University of Illinois at Urbana-Champaign, Urbana, Illinois 61801, USA}
\affiliation{Research Center for Astronomy and Applied Mathematics, Academy of Athens, Athens 11527, Greece}

\author[ 0009-0002-7498-6899]{Priyesh Kumar Tripathi}
\affiliation{Aryabhatta Research Institute of Observational Sciences (ARIES) \\
Manora Peak  \\
Nainital 263001, India  }
\begin{abstract}
We perform the numerical simulations of axisymmetric, relativistic, optically thin jets under the influence of the radiation field of an accretion disk. We show that starting from a very low injection velocity at the base, jets can be accelerated to relativistic terminal speeds when traveling through the radiation field. The jet gains momentum through the interaction with the radiation field. We use a relativistic equation of state for multi-species plasma, which self-consistently calculates the adiabatic index for the jet material. All the jet solutions obtained are transonic in nature. In addition to the acceleration of the jet to relativistic speeds, our results show that the radiation field also acts as a collimating agent. The jets remain well collimated under the effect of radiation pressure. We also show that if the jet starts with a rotational velocity, the radiation field will reduce the angular momentum of the jet beam.      
\end{abstract}


%%%%%%%%%%%%%%%%%%%%%%%%%%%%%%%%%%%%%%%%%%%%%%%%%%

%%%%%%%%%%%%%%%%% BODY OF PAPER %%%%%%%%%%%%%%%%%%

\section{Introduction}
Collimated plasma outflows, commonly referred to as astrophysical jets, that show elongated morphology are observed in a variety of galactic and extragalactic sources, including young stellar objects (YSOs), neutron stars, gamma-ray bursts (GRBs), microquasars and active galactic nuclei (AGNs). The bulk velocity of plasma in these jet beams is quite often estimated to reach values close to the speed of light. The occurrence of jets seems to be prevalent in any scenario where there is mass accretion onto a central object. Interestingly, the simultaneous radio and X-ray observations have revealed a significant correlation between the different spectral states of accretion disks and jet states in microquasars, as demonstrated in studies by \cite{fgr10,rsfp10}. This correlation suggests that the jet indeed originates from the accretion disk. Furthermore, the jet is launched from the inner part of the accretion disk, within 100 Schwarzschild radii ($\rg$) \citep{jbl99}.  Additionally, since the jet originates from a region close to the black hole horizon, it will pass through the radiation field of the disk and will undoubtedly be influenced by it. The equations of radiation hydrodynamics govern the evolution of the ionized jet plasma interacting with the radiation field. Radiation hydrodynamics is a well-established field with contributions from several authors \citep{mm84,k98,p06}. These equations have been extensively used in the steady-state analytical investigations of radiatively driven jets and winds. \cite{i80} examined the flow of particles above an alpha or standard disk \citep{ss73}. His initial investigation neglected the influence of radiation drag. However, later in his significant paper \citep{i89}, radiation drag was included, setting an upper limit on the terminal speed of plasma in the radiation field above an infinitely extended alpha disk. This limit, termed the ``magic speed'', is approximately $0.45$ times the speed of light. 
\cite{sw81} studied the acceleration and collimation of the particle beams in the funnel region of thick accretion disks. The jets composed of pair plasma (electron-positron plasma) were accelerated to terminal Lorentz factors of $\sim 3$. Using full special relativistic treatment \cite{f96} examined the accretion disk winds and showed the terminal speeds for the winds above a standard accretion disk are lower than the magic speeds of \cite{i89}. Alongside its main findings, this investigation introduced an interesting question about the feasibility of the radiation field's role in collimating the jet. The winds acquired angular momentum from the radiation field, resulting in less efficient collimation. To address the collimation of jets through the radiation field, \cite{f99} studied the jets confined by the disk corona, and the jet flow displayed a notably high degree of collimation. {In a certain range of flow parameters, the inner part of the accretion disk may harbor shock waves in advective type disk models as shown by \cite{f87,c89}. The post-shock region, being hotter and denser than the pre-shock disk, has been proposed as the illusive corona since it emits mostly as hard power-law radiation \citep{ct95}.} Therefore, it is possible that the compact corona is responsible for the jet activity. Numerical simulations of advective disk models have shown that the bipolar outflows are automatically generated from the inner post-shock disk (PSD) due to the presence of an additional thermal gradient \citep{sm94,msc96,mrc96,dcnm14,lcksr16,kgbc17}.  In the disk model proposed by \cite{ct95}, a standard disk is surrounded by a sub-Keplerian disk (called SKD), which can undergo a shock transition and create a hot and compact puffed-up PSD. A more detailed and consistent method was employed in the two-temperature advective accretion flow to compute the emissivity and spectra from these flows
\citep{scl20,sc22,sscl23}.
Chattopadhyay and Chakrabarti \citep{cc00a,cc00b, cc02} investigated the effect of intense radiation of PSD on the outflowing jets and showed that they can achieve a velocity in the range of $0.2c-0.3c$ ($c$ is the speed of light) through radiative acceleration. Later, in the relativistic hydrodynamic regime \cite{c05} considered the radiation field of the outer Keplerian disk along with the PSD radiation field and showed that the radiation field of the PSD is the primary accelerating agent that can accelerate the pair-plasma jets to Lorentz factor $\gamma>2$, while the radiation generated by the Keplerian disk collimates the jet. In the general relativistic radiation hydrodynamic regime, \citep{vc18, vc19}  have demonstrated that including the radiation field of SKD enhances the acceleration of jets and lepton-dominated jets reach up to $\gamma\sim10$. Contrary to the popularly accepted belief that radiation can not accelerate the jet to relativistic speed \citep{ggmm02}, a noteworthy point highlighted from these studies is that radiation can accelerate jets to relativistic terminal speeds. We will discuss later why radiation from one set of accretion disk models couldn't accelerate jets to relativistic speeds while others did. \\

 The radiatively driven jets can be examined in two-dimensional studies using the semi-analytical approach \citep{fth01,c05} and numerical simulations. The analytical studies of rotating jets assumed the gas pressure to be negligible compared to the radiation pressure. Without this assumption, it is impossible to compute the streamlines of the jet in the analytical studies. In multi-dimensional studies, numerical simulations of relativistic jets have become an essential tool in recent years for understanding its complicated physics \citep{m19,skrhc21}. Despite the considerable advancements through analytical investigations, there has been a limited number of numerical simulations focusing on radiatively driven outflows and jets. \cite{eg85} performed a two-dimensional radiation-coupled, Newtonian hydrodynamic simulation of super-Eddington accretion and showed the self-consistent formation of the axial jet with velocity $\sim 0.3c$. \cite{ommk09} suggested a unified model for different spectral states (low/hard state, high/soft state) based on global, two-dimensional radiation-magnetohydrodynamic simulations and showed the presence of disk outflows driven either by radiation pressure force or magnetic pressure force. In addition to acceleration, these simulations also show that the geometrically thick disks help collimate the jet. The radiatively driven model is one of the proposed models, which explains the generation of astrophysical outflows from luminous accretion disk sources. Hence, the generation of outflows does not depend on black hole spin \citep{sn15}.
 
 For the advective type disks, \cite{cc02a} studied the acceleration of jets using a Smooth Particle Hydrodynamic (SPH) code and showed that radiative acceleration produces supersonic outflows. \cite{jcl22} simulated radiatively driven fluid jets around a non-rotating black hole. The study focused on the influence of radiation originating from the inner compact corona and outer sub-Keplerian part of the accretion disk on the jet acceleration in the non-relativistic radiation hydrodynamic regime. As the jets have relativistic speeds, one must study the jet in the relativistic framework. Hence, in \cite{jdc22}, a numerical simulation of a conical jet was performed along the axis of symmetry to explore the behavior of radiatively driven, time-dependent, relativistic jets around black holes. We found out that radiative acceleration can have a significant impact on the jet propagation velocity to the extent that pair-dominated jets can be accelerated to ultra-relativistic terminal speeds. These authors also showed that a time-dependent radiation field produced by an oscillating disk can produce a series of shocks in jets. This simulation code was based on special relativistic TVD (total variation diminishing) routine \citep{rcc06,crj13,jc23} with its thermodynamics described by an approximate relativistic equation of state \citep{cr09}, while gravity was introduced through the weak field approximation. In this work, we extend the investigation to two dimensions by performing simulations of axisymmetric jets in cylindrical geometry. Numerical simulations performed to investigate the morphology of relativistic jets usually start with the supersonic initial conditions \citep{mmi97,jc23,skr21}. In this study, we are interested in the acceleration of the jet, which originates very close to the black hole. Hence, we start our simulations from subsonic injection speeds. In contrast to other simulations, for example,  performed by \cite{snpz13,sn15,to15,uota22}, we do not probe the jet generation by the radiation pressure in this work. The jet originates because of accretion shock, as is the case for transonic accretion disks \citep{msc96,mrc96,dcnm14,lcksr16}. And we assume the jet has been launched with subsonic speeds at some base height. The main focus of this study is to highlight the evolution of the jet as it goes through various phases, starting from the denser subsonic phase to the lighter supersonic phase. Additionally, we must also emphasize that the disk model taken in our study is completely different from the disk model adopted by the previously mentioned authors. We are not trying to radiatively launch jets from the accretion torus. The accretion torus has very little
 advection; as a result, the density of the torus is high. So, the jets being launched are also not optically thin, which results in multiple scattering of the photons with matter. Our disk model is the advective disk model. The maximum mass density in such a disk is a few orders of magnitude lower. As jets are launched due to the action of the accretion shock, only a fraction of accreting matter is ejected. Also, the very process of ejection produces a divergent flow geometry! So, the density of outflows from such a disk is low. In this paper, we have investigated the radiative acceleration of an optically thin jet in the presence of a radiation field produced by an advective disk. Apart from the jet density, the nature of radiative moments strongly depends on the disk models adopted. It has been shown in many analytical investigations \citep{f05,vkmc15,vc17} that the advective disks can accelerate the jets up to speed greater than $0.9c$. In this paper, we explore how the radiation field of the advective accretion disk influences the dynamics and morphology of the jet, with the help of numerical simulations.  
 
 The radiation field of the disk accelerates the jet to supersonic speeds. As the jet is transonic in nature, the use of a fixed polytropic index equation of state (EoS) can lead to inconsistent results. Hence, it is recommended to utilize an EoS that considers the self-consistent development of the adiabatic index \citep{dho96}.  
 This paper examines the dynamics of radiatively driven jets in general, with a specific focus on the impact of the efficiency of radiative acceleration of a laterally expanding jet. The paper is structured as follows: In Section \ref{sec:equations}, we provide an overview of the governing equations and details of the simulation setup are given in Section \ref{sec:setup}. Section \ref{sec:results} presents the analysis and findings of this work. Finally, Section \ref{sec:concl} offers a brief discussion and conclusion of the research.     

\section{Equations and Numerical Method}
\label{sec:equations}
We perform simulations of axisymmetric optically thin relativistic jets under the influence of the radiation field of the accretion disk in the cylindrical geometry ($r,\theta,z$). The metric is given by

\begin{equation}
ds^2=-(1+2\Phi)c^2dt^2+dr^2+r^2d\theta^2+dz^2; \,\,\,\, \Phi=\frac{GM}{R-\rg}
\label{eq:metric_potn}
\end{equation}
Here, $R=\sqrt{r^2+z^2}$  is the radial distance from the central object, $r,\theta,z$ are usual coordinates in cylindrical geometry, and $\rg=2GM/c^2$ is the Schwarzschild radius. The major advantage of using this weak field approximation in the metric is that gravity enters only in the time component of the metric, keeping the space flat. Therefore, one can avoid the computation of curved photon propagation path and yet
get the effect of the horizon and strong gravity.
For jets, this is fine as jets have very long length scales. Moreover, such an arrangement maintains the tensorial property of the equations of motion, \\
The detailed derivation for the full set of equations of radiation hydrodynamics can be found in various literature\citep{mm84,k98,p06,r07}. Here, we present only the conservative form of these equations:

\begin{subequations}
\begin{equation}
\frac{\partial D}{\partial t}+\frac{1}{r}\frac{\partial}{\partial r} \left[r\alpha Dv^r\right]+\frac{\partial}{\partial z}\left[\alpha Dv^z\right]=0    
\label{eq:continuity}
\end{equation}

\begin{equation}
\frac{\partial M^r}{\partial t}+\frac{1}{r}\frac{\partial}{\partial r}  \left[r\alpha\left(M^rv^r+p\right)\right]+\frac{\partial}{\partial z}\left[\alpha M^rv^z\right]=\frac{\alpha p}{r}+\frac{\alpha M^\theta v^{\theta}}{r}-E\frac{\partial \alpha}{\partial r}+G^r    
\label{eq:momentum_r}
\end{equation}

\begin{equation}
\frac{\partial M^\theta}{\partial t}+\frac{1}{r}\frac{\partial}{\partial r}\left[r\alpha M^\theta v^r\right]+\frac{\partial}{\partial z}\left[\alpha M^\theta v^z\right]=-\frac{\alpha M^\theta v^r}{r}+G^\theta    
\label{eq:momentum_theta}
\end{equation}

\begin{equation}
\frac{\partial M^z}{\partial t}+\frac{1}{r}\frac{\partial}{\partial r}\left[ r\alpha M^zv^r\right]+\frac{\partial}{\partial z}\left[\alpha (M^zv^z+p)\right]=-E\frac{\partial \alpha}{\partial z}+G^z    
\label{eq:momentum_z}
\end{equation}

\begin{equation}
\frac{\partial E}{\partial t}+\frac{1}{r}\frac{\partial}{\partial r} \left[r\alpha (E+p)v^r\right]+\frac{\partial}{\partial z}\left[\alpha (E+p)v^z\right]=-M^r\frac{\partial\alpha}{\partial r}-M^z\frac{\partial\alpha}{\partial z}+G^t    
\label{eq:energy}
\end{equation}
\end{subequations}
In equations, \ref{eq:continuity}-\ref{eq:energy}, $D,\,M^r,\,M^z,\,E$ represent the conserved fluid quantities, namely the mass density, radial and axial component of momentum density, and the total energy density of the fluid, respectively. The components of the 3-velocity are $v^r,\,v^\theta,\,v^z$, and  $\alpha=\sqrt{(1+2\Phi)}$. It may be noted that the time and space components of 4-velocities are given as,
$u^t=\gamma/\sqrt{\alpha}$, $u^i=\gamma v^i/\sqrt{g_{ii}}$, respectively, here $g_{ii}$ are the metric tensor components and $\gamma=1/\sqrt{(1-v_iv^i)}$ is the Lorentz factor. $G^\mu\,(\mu=t,r,\theta,z)$ are the components of radiation four-force density given as 

\begin{subequations}
\begin{equation}
G^r=G^{\hat{r}}_{co}+\frac{\gamma-1}{v^2}v^rv_iG^{\hat{i}}_{co}
\label{eq:gr}    
\end{equation}
\begin{equation}
G^\theta=\frac{1}{r}\left[G^{\hat{\theta}}_{co}+\frac{\gamma-1}{v^2}v^\theta v_iG^{\hat{i}}_{co}\right]
\label{eq:gtheta}    
\end{equation}
\begin{equation}
G^z=G^{\hat{z}}_{co}+\frac{\gamma-1}{v^2}v^zv_iG^{\hat{i}}_{co}
\label{eq:gz}    
\end{equation}

\begin{equation}
G^t=\frac{\gamma}{\alpha}v_i G^{\hat{i}}_{co}    
\label{eq:gt}
\end{equation}

\begin{equation}
G^{\hat{i}}_{co}={\rho_e} F^i   \,\,\,\,\,\, (i=r,\theta,z)
\label{eq:gico}
\end{equation}

Where $F^i$ is given as 
\begin{align}
F^i &=-\gamma^2v^i\mathcal{E}+\gamma\left[\delta^i_j+\left(\gamma+\frac{\gamma^2}{\gamma+1}\right)v^iv_j\right]\mathcal{F}^j \nonumber \\
&-\gamma v_j\left[\delta^i_k+\frac{\gamma^2}{\gamma+1}v^iv_k\right]\mathcal{P}^{jk}    
\label{eq:Fico}
\end{align}   
\end{subequations}

In equation \ref{eq:Fico}, $\,\mathcal{E}=\frac{\sigmat}{\mel c}E_{\rm rd},\,\mathcal{F}^i=\frac{\sigmat}{\mel c}{F}_{\rm rd}^i,\,\mathcal{P}^{jk}=\frac{\sigmat}{\mel c}{P}_{\rm rd}^{jk}$. Here, $E_{\rm rd},\,F_{\rm rd}^i,\,P_{\rm rd}^{ij}$, represent the radiation energy density, three components of radiation flux, and six independent components of radiation pressure \citep{c05}. $\rho_e$ is the mass density of leptons and $\mel$ is the mass of electron. $\sigmat$ is the Thomson scattering cross-section.
In Appendix \ref{app:radmom}, expressions of the radiative moments are shown in equations \ref{eq:eng_rad}, \ref{eq:forc_rad} \& \ref{eq:pres_rad}.

Additionally, we use an approximate relativistic equation of state (abbreviated as CR EoS) proposed by \cite{cr09} as a closure relation for the set of equations \ref{eq:continuity}-\ref{eq:energy}. The algebraic form of CR EoS is given as   
\begin{equation}
e=\rho f
\label{eq:eos}
\end{equation}

where,
\begin{equation}
f=1+(2-\xi)\Theta\left[\frac{9\Theta+6/\tau}{6\Theta+8/\tau}\right]+\xi\Theta\left[\frac{9\Theta+6/\eta\tau}{6\Theta+8/\eta\tau}\right]
\label{eq:eos2}
\end{equation}
In equations (\ref{eq:eos}, \ref{eq:eos2}) $\rho$ represents the rest mass density of fluid, $e$ is the energy density of the fluid, $\xi=\np/\nel$ is the proton fraction, and $\eta=\mel/\mp$ 
$\nel$, $\np$, and $\mp$ are the electron and proton number
density,  the rest mass of the proton. $\Theta=p/(\rho~c^2)$ is a measure of temperature and $\tau=2-\xi+\xi/\eta$.  We solve the equations \ref{eq:continuity}-\ref{eq:energy} numerically using a relativistic TVD simulation code \citep{rcc06,crj13,jc23}. The details of the code are also given in \citep{jdc22}. 

\begin{figure*}
	\includegraphics[width=18cm,height=10cm]{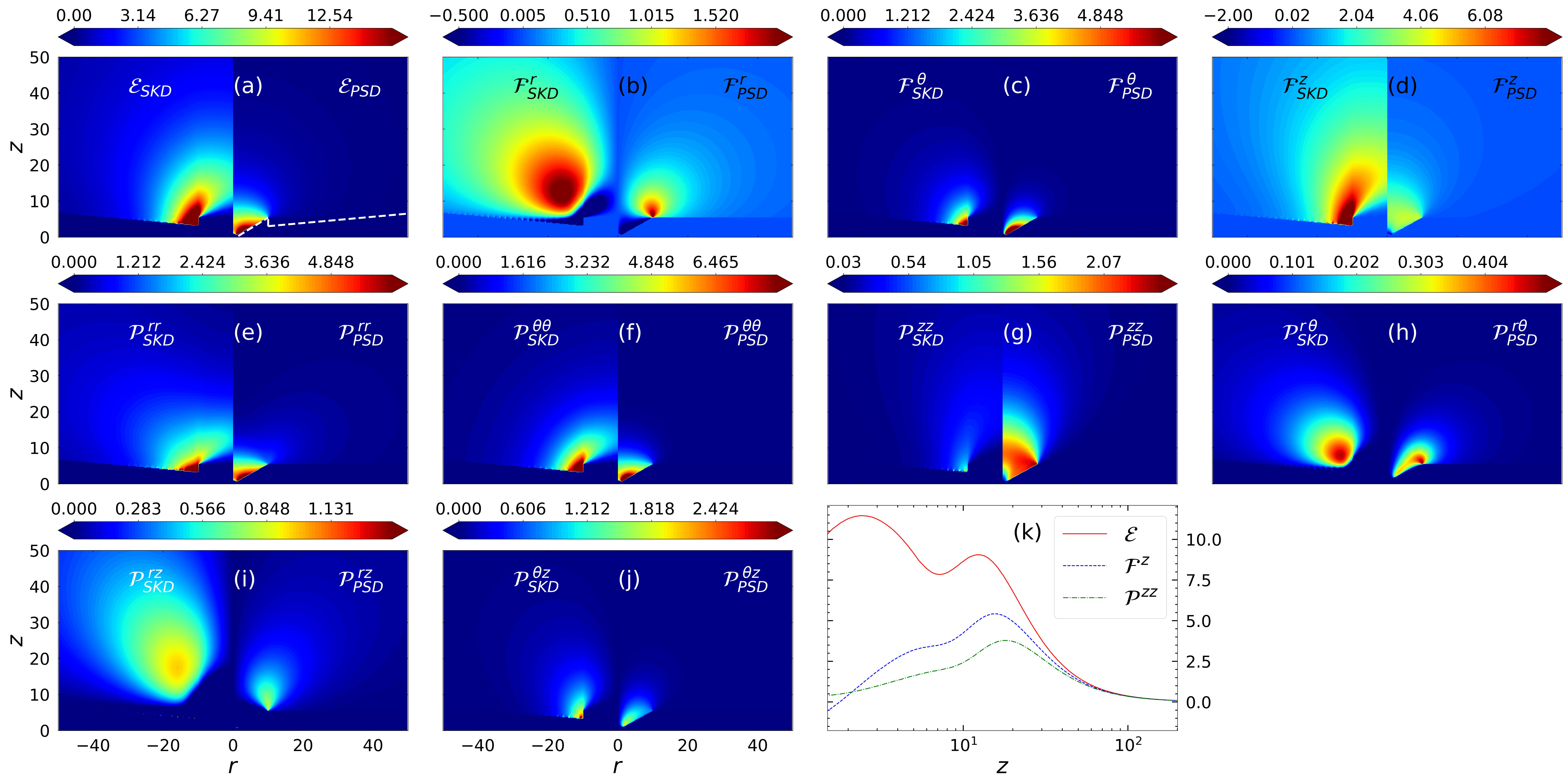}
    \caption{The distribution of radiative moments (a) $\mathcal{E}$, (b - d) $\mathcal{F}^{i}$,
    and (e - j) $\mathcal{P}^{ij}$ for the SKD (left half of each panel) PSD (right half); (k) Radiative moments along the z-axis. The dashed white line in panel (a) represents the surface of the disk; radiative moments are computed above it.}
    \label{fig:psd_moms}
\end{figure*}

\section{Simulation Setup}
\label{sec:setup}
To discretize the computational domain, we utilize a uniform spacing grid with a constant spacing of $dr=dz=0.2\rg$. We adopt the unit system where $2G=M=c=1$. In such a unit system, the gravitation potential (equation \ref{eq:metric_potn}) becomes
$$
\Phi=\frac{0.5}{R-1}
$$

The time in the simulation code is in units of $\tg=\rg/c$. The computational grid spans $500\times 10000$ cells, covering a size of $100\rg \times 2000\rg$. The $z$-axis is bounded by a reflection boundary condition, while the outer $r$ and $z$ boundaries are treated as outflow boundaries. As mentioned earlier in this work, we do not focus on the generation of the jet, we assume that the jet has originated from the disk itself with some initial jet base parameters (velocity, density, and pressure), so we inject the jet material using a nozzle of radius $5\rg$, and the initial length of the beam is also taken to be $5\rg$. The jet material is continually injected from a fixed jet base. 
The plasma composition of jets is taken to be $\xi=0.001$. The injection parameters are taken to be $\rho_j=1000,\,v^z_j=0.001$ at $z=3.0$. The pressure in the jet beam is $p_j=100$. These injection parameters are taken from a one-dimensional cylindrical radiatively driven jet that has a sonic point at $z=5.0$. The ambient medium is static and $1000$ times lighter than the jet ($\rho_a=1.0$), and the pressure of the ambient medium is $p_a=p_j/10.0$. We have considered disk-luminosity values obtained for an accretion disk around a black hole of $10{\rm M_{\odot}}$. 
The radiation field is computed from the radiative emissivity computed from the accretion rate of the matter, although the accretion disk is auxiliary in this simulation.

Physically, it implies that it is about three orders of magnitude lower than the maximum mass density of the accretion disk on the equatorial plane. For example, 
the maximum value of mass density of the accretion disk on the equatorial plane for a non-rotating black hole for accretion rate $\sim 10 \dot{M}_{Ed}$, is of the order of $10^{-6} \mbox{g cm}^{-3}$. The jet base is located at some distance above the equatorial plane, and the outflow/jet has an expanding flow geometry. Therefore, the number density of the jet should be 2-3 orders of magnitude lower than that on the equatorial plane of the accretion disk. In addition to that, our focus is to study the lepton-dominated jets.  
Based on these assumptions, the number density at the jet base is three orders of magnitude less the maximum central number density of the disk and considered $n_{\rm jetbase}\sim 5.98 \times 10^{14}$cm$^{-3}$, and $\rho=1$ would correspond to $\sim 10^{-15}$g cm$^{-3}$. 

\section{Analysis and Results}
\label{sec:results}

\subsection{Nature of the Radiative Moments}
\label{sec:moments}

We calculate the distribution of radiative moments by following the method outlined in \cite{c05} for advective-type disks \citep{f87,c89,cc11,scl20}. The disk consists of two components, namely, an inner hot puffed-up post-shock disk (PSD), which is a supply of hot electrons like a corona, and an outer sub-Keplerian disk (SKD). These disks have low to moderately high radiative efficiency \citep{scl20,sc22}. The outer edge of PSD and the luminosity are calculated using the algebraic expressions given by \cite[][also see equation \ref{eq:xs_mdot}]{vkmc15}. In Fig. \ref{fig:psd_moms}, we have plotted the distribution of various radiative moments in panels (a)-(j). The exact form of the radiative moments is given in Appendix \ref{app:radmom}. Since the attenuation of the radiative intensity is marginal, because the jet is optically thin, therefore, the intensity falls off geometrically from the disk surface only and it is not reduced by attenuation. In Fig. \ref{fig:psd_moms}a, right panel, the dashed white line shows the accretion disk surface. The left half of each panel shows the distribution of moments due to SKD, and moments due to PSD are plotted in the right half. The accretion rate of the SKD controls the shock position and the luminosities of the PSD and SKD. The luminosities of the disk components are calculated using the algebraic functions given in \cite{vkmc15}, which are based on the general radiative transfer model of \cite{mc08}.  
For $\dot{m}_{\rm sk}=8.5$, the shock location or outer edge of PSD is at $\xs=10.0$. The luminosity of PSD is $\lps=0.25 L_{\rm Edd}$, and the luminosity of SKD is $\lsk=0.04 L_{\rm Edd}$, where $L_{\rm Edd}$ is the Eddington luminosity ($L_{\rm Edd}=1.26\times 10^{38} M/M_{\odot}{\rm erg\,s^{-1}}$). The distribution of these moments is highly anisotropic inside the funnel region of the disk. The radiation energy density is dominant in comparison to other moments. $\mathcal{F}^r$ is negative inside the funnel; hence it should push the material towards the jet axis. Panel \ref{fig:psd_moms}-(d) shows that $\mathcal{F}^z$ is positive inside the funnel. Hence, it will accelerate the jet, but $\mathcal{P}^{zz}$ generates the radiation drag and tries to slow down the jet material.  
One can clearly see that inside the funnel $\mathcal{F}^z>\mathcal{F}^\theta>\mathcal{F}^r$, while 
$\mathcal{P}^{\theta\theta}>\mathcal{P}^{rr}>\mathcal{P}^{zz}$. This implies that the net radiative acceleration will be highest along the $z$ direction. $\mathcal{F}^\theta$ and $\mathcal{P}^{\theta i}$ have non-zero values. So as $\mathcal{F}^\theta$ would like to spin up the jet,
$\mathcal{P}^{\theta i}$ will tend to induce radiation drag in Equation \ref{eq:momentum_theta}. The drag along the $\theta$ direction is larger than the acceleration term; hence $v^\theta$ will be reduced eventually.
In addition to the magnitude of radiation energy density and pressure, the radiation drag also depends on the propagation velocity. Inside the funnel region of the disk, the propagation velocities are small. Hence, the drag is not very significant. One must also notice that the radiation from the SKD is blocked due to the shadow effect of the PSD \citep{c05}, so inside the funnel region, the dynamics of the jet is only governed by the radiation field of the PSD only because the jet can not see the radiation field due to the most luminous part of the SKD. In panel (k), we have plotted the distribution of $\mathcal{E},\,\mathcal{F}^{z},\,\mathcal{P}^{zz}$ along the jet-axis. The distribution of all these moments reveals two prominent peaks, a result of the distinct maxima in moments from SKD and PSD at different positions. 
This type of distribution of moments is one key aspect of the disk model we have adopted. This helps in the multi-stage acceleration of the jets. Notably, for disk dimensions mentioned above, at distance $z>100r_g$ it becomes evident from Fig. \ref{fig:psd_moms}(k) that $\mathcal{E}\sim\mathcal{F}^{z}\sim\mathcal{P}^{zz}$, implying that the radiation field closely resembles that produced by a point source. A closer inspection into the expression of the radiation force (Eq. \ref{eq:Fico}) shows that the energy density and pressure terms are combined with odd powers of velocity components and also come with a negative sign. These terms are called the radiative drag terms. Meanwhile, the flux is associated with the quadratic power of $v_i$s and comes with a positive sign. Solving for $F^i=0$, one will get a measure of velocity called the equilibrium speed ($v_{\rm eq}$), above which radiative deceleration sets in. It has been shown before that for a source, for which ${\cal E} \sim {\cal F}^z \sim {\cal P}^{zz}$, $v_{\rm eq} \rightarrow 1$ \citep{cdc04}. It implies that when $v_{\rm eq} \rightarrow 1$, there is actually no upper limit of speed. Therefore, if the brightness of the source is high and yet ${\cal E} \sim {\cal F}^z \sim {\cal P}^{zz}$, then, the jet can be accelerated to ultra-relativistic speed.

\subsection{Radiative Acceleration of the jets: Model-A}
\label{sec:results1}
\begin{figure*}
	\includegraphics[width=18cm,height=7cm]{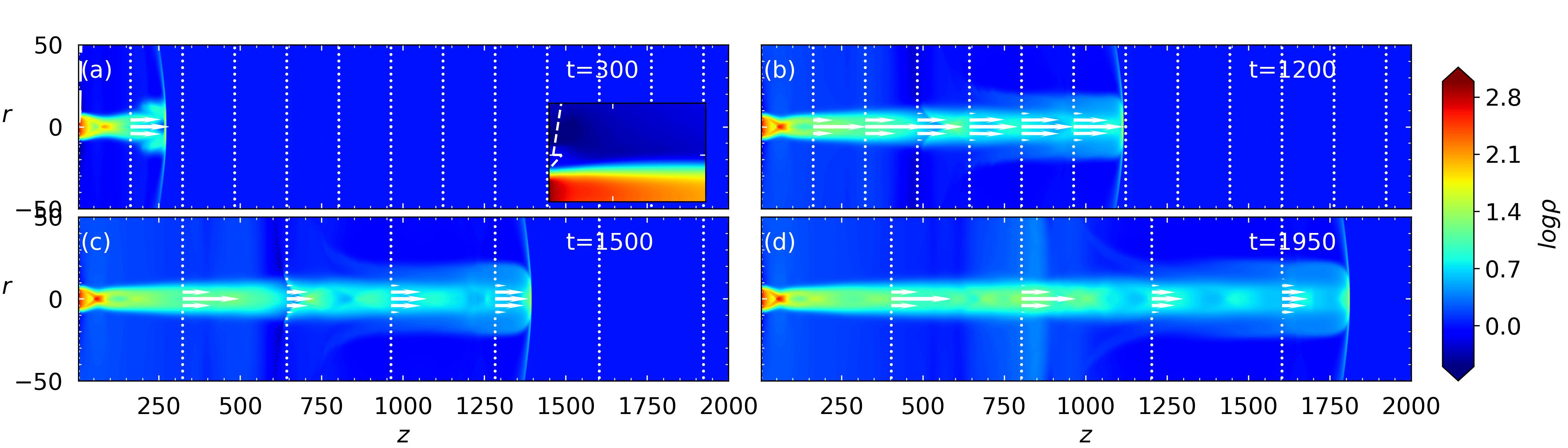}
    \caption{The density contours along with vector field of momentum flux ($M^r, M^z$) for the Model-A. \bf{The photosphere is marked by the dashed white lines in the zoomed inset of panel (a)}.}
    \label{fig:jetlofac}
\end{figure*}

\begin{figure*}
	\includegraphics[width=18cm,height=7cm]{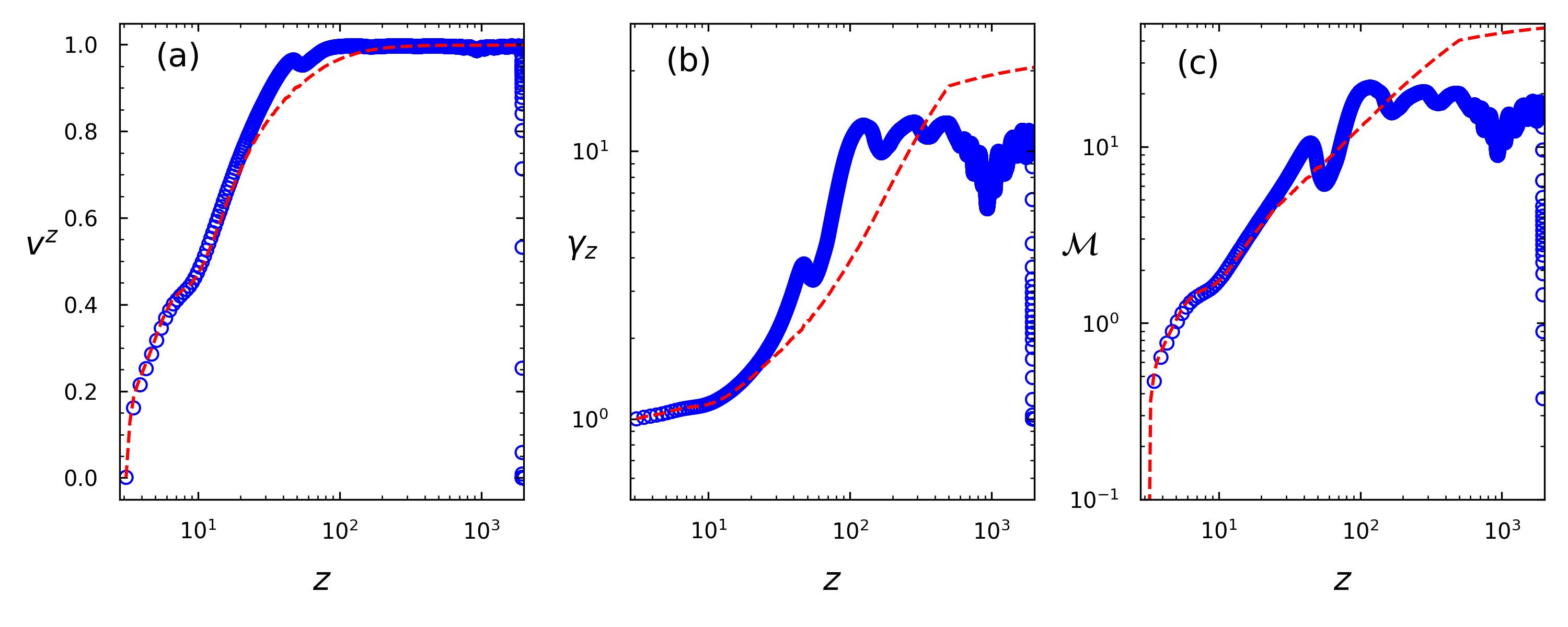}
    \caption{Variation of $v^z,\,\gamma_z,\,$and relativistic Mach number$\,\mathcal{M}$ is plotted in panel (a) to (c). The dashed red line represents the solution for a steady-state jet, and blue open circles correspond to the variables along the jet spine of the 2D simulation. }
    \label{fig:spineplot}
\end{figure*}

The jets originate from a region very close to the black hole, and the jet base velocity is expected to be very small. Also, the jet base temperatures are fairly high, so the local sound speed is high. Hence, the jet must be subsonic in regions close to the central object. As the internal thermal energy and the radiation drive the jet to a higher velocity, the jet starts to expand, making it less hot. Hence, at larger distances, the jet must be supersonic. In our previous study \citep{jdc22}, we have already shown that the radiation field can efficiently accelerate the jet to a relativistic terminal speed by performing the one-dimensional conical jet simulation. In multidimensional simulations, additional features like backflow, formation of the cocoon, and fluid instabilities also form as the jet propagates in the medium. Moreover, the radial and azimuthal components of flux and various components of radiative moments become dynamically important. In order to check how good was our previous one-dimensional simulation, first, we run a two-dimensional simulation of the jet propagation by considering the contribution of only those radiative moments that are dynamically important in a one-dimensional study along the direction of jet propagation, namely $\mathcal{E},\,\mathcal{F}^z,\,\mathcal{P}^{zz}$.  In Fig. \ref{fig:jetlofac}, we have plotted the density contours along with the vector field of momentum flux ($M^r,\,M^z$) to show the morphology of the jet at different time epochs. We have also marked the accretion disk surface through dashed white lines in the zoomed-inset of the panel (a). As seen for panel (a), initially, the jet is very heavy in comparison to the ambient medium. As the jet propagates, it gains momentum and also starts to expand. As a result, the density inside the jet beam decreases. In panels (c) and (d), the low-density cavities can also be seen in the late phases of evolution. As the jet keeps on propagating through the ambient medium, additional features along the jet beam also start to form.  
As the jet is being accelerated under the radiation field only along the z direction, it remains well-collimated.
In Fig. \ref{fig:spineplot} (a-c), we have compared the $z$ component of three velocity i. e., $v^z$ and the Lorentz factor related to $v^z$ and the relativistic Mach number ($\mathcal{M}={\gamma_zv^z}/{(\gamma_{\rm cs}c_s)}$),
respectively, between purely one-dimensional simulation as in \cite{jdc22} with that along the axis of symmetry ($0,z$) of the present simulation
(Fig. \ref{fig:jetlofac}). The dashed red line corresponds to a one-dimensional jet \citep{jdc22}, and the blue open circles show the profiles of flow variables of the axisymmetric two-dimensional simulations along the jet axis of Fig. \ref{fig:jetlofac}. 
One can clearly see that the velocity near the jet base is very small, but it eventually increases through the radiative acceleration. The variation of the Lorentz factor clearly shows that the jet is accelerated to terminal relativistic speeds. The Mach number plot indicates that the jet is transonic in nature, it starts with a subsonic inner boundary condition and becomes supersonic at $z \sim 5$. 
Both the solutions in Fig. \ref{fig:spineplot}, are in fair agreement up to a distance of $20\rg$ because, in the initial phases, the internal structures do not form in the jet beam in multi-dimensional simulations. At the intermediate distances ($30 \rg \lsim z \lsim 300 \rg$), the 2D jet has a higher velocity than 1D. The jet area expansion rate differs from the one-dimensional geometry in multi-dimensional simulations. Hence the thermal expansion can give an additional boost to the velocity. At larger distances, as the jet becomes highly supersonic and interacts with the ambient medium, it generates multiple additional internal structures, and the jet moves with a lower speed in comparison to the one-dimensional estimate. The difference in velocity is enhanced at the regions where the jet beam shows additional structures like density-enhanced regions and expansion fans. In a one-dimensional study, the only dynamically important velocity component is along the direction of propagation, but in multi-dimensional simulations, the lateral component of the velocity also develops as the jet starts to push the medium in front of it. Hence, at larger distances, the terminal Lorentz factor $\gamma_z \sim 20$ of one-dimensional simulation does not match with a multi-dimensional simulation where the terminal $\gamma_z \sim 11$.  

\begin{figure*}
	\includegraphics[width=18cm,height=8cm]{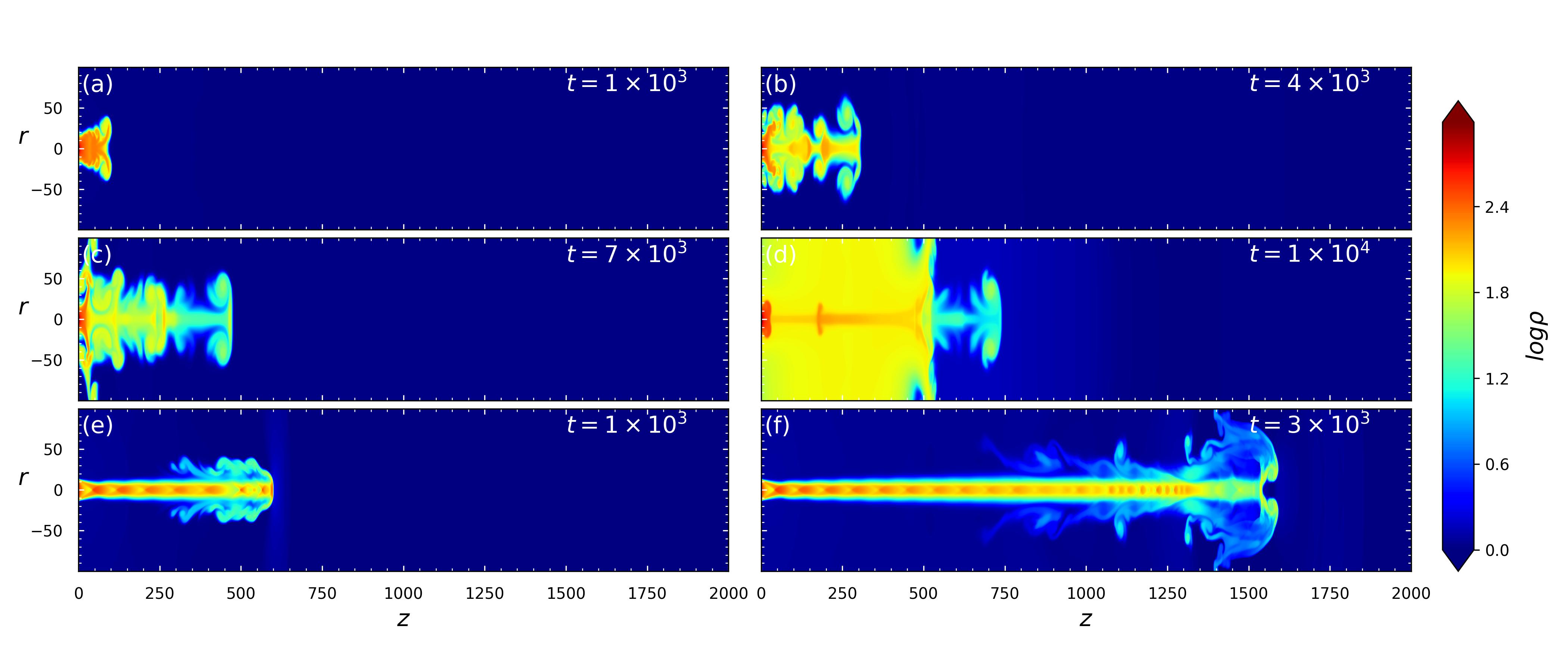}
    \caption{The density contours for a thermally driven jet at different time steps as mentioned in panels (a)-(d). The density contours for the jet-driven with $10\%$ of the radiation field are plotted in panels (e)-(f).}
    \label{fig:jetthermal}
\end{figure*}

As the jet expands, it gains momentum at the expense of internal energy. Hence, thermal driving can also accelerate the jets. To highlight the efficiency of radiative acceleration, we have also compared the evolution of the thermally driven jet beam without the influence of the radiation field of the accretion disk
with the one that is driven by the $10\%$ of the radiation field. In Fig. \ref{fig:jetthermal}a-d, we plotted the density contours for a thermally driven
at different time step $t=10^3,~4\times 10^3,~7\times 10^3$ and $10^4$, respectively. One can see that the thermal energy of the initial jet beam is sufficient to drive the jet material against the gravity of the central object but not enough to drive the outflow to ultra-relativistic speeds, as the time stamps in different panels indicate that after a significant amount of time, the propagation of the injected material in the lateral direction is comparable to its propagation along the jet axis.  
The vortex-like features around the jet head start to form very early in thermally driven jets. The instability of the thermally driven jet leads to its eventual dissipation as the slow and hot jets are known to be unstable \citep{rz13}. This instability is primarily caused by the continuous disruption of the mixing layer that grows towards the jet axis, which can be seen from the panels Fig. \ref{fig:jetthermal}(a)-(d). 
{In order to check the effect of radiation on the stability of the jet, we also perform a simulation with only $10\%$ of the radiation field. These results are plotted in panels (e)-(f). The evolution in this configuration shows that even with a very small amount of radiation, the jet propagates with a collimated geometry, and the growth of the mixing layer reduces, which makes the jet stable against the instabilities that disrupt the jet in a purely thermal-driven case.
The time stamp on panels Fig. \ref{fig:jetthermal} e \& f shows that the jet head has traveled more than twice the distance in less than one-third of the time if the jet is
driven by only one-tenth the luminosity of the disk.}    
Hence, we can conclude that in our simulation, radiation is the primary driving agent which accelerates the jet.

\section{Morphology of jet with all radiative moments: Model-B}
\label{sec:allmoms}

\begin{figure*}
	\includegraphics[width=19cm,height=8cm]{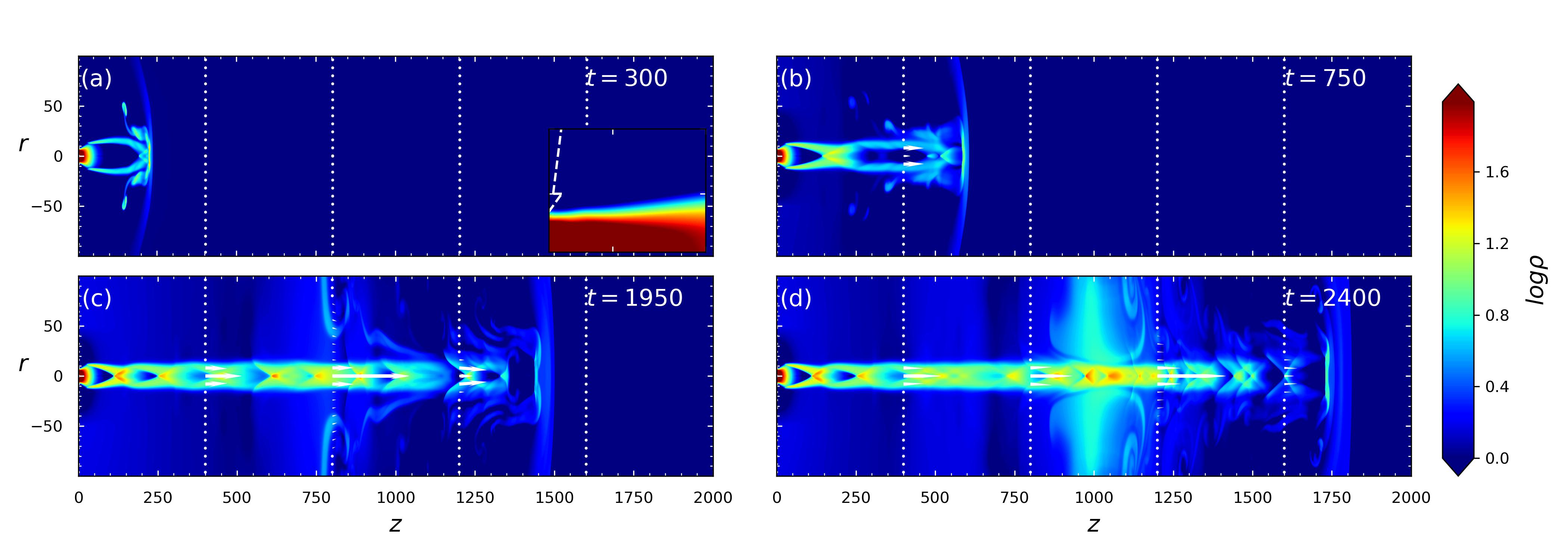}
    \caption{Contours of density plotted at different time steps for the jet with all components of radiative moments (Model-B).}
    \label{fig:allmoms_dens}
\end{figure*}

\begin{figure*}
	\includegraphics[width=18cm,height=7cm]{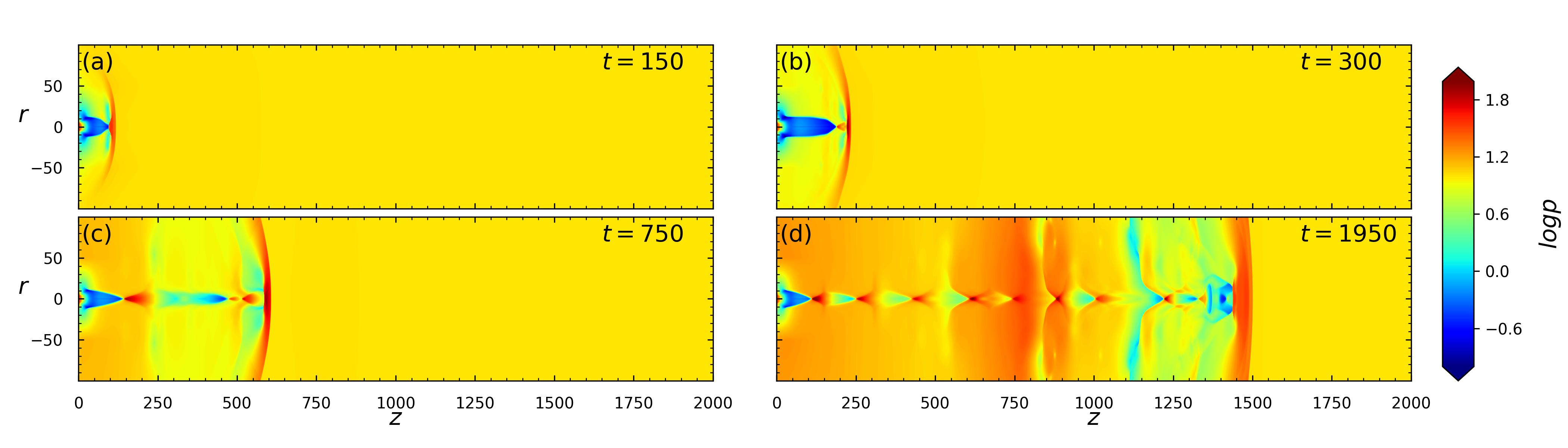}
    \caption{Contours of pressure for Model B. The time steps for the snapshots are mentioned in the respective panel.}
    \label{fig:pres}
\end{figure*}

In a multi-dimensional simulation, all ten components of the radiative moments are dynamically important. It is clearly evident from Eq.\ref{eq:momentum_r}-\ref{eq:momentum_z} that including all the moments will influence the other velocity components in addition to the velocity along the direction of propagation of the jet, which can result in additional differences in the morphology of the jet. In Fig. \ref{fig:allmoms_dens}, we have plotted the contours of the density for the jet under the influence of all the radiative moments at different times. As we have taken the contribution of radiative flux along the $r$ and $\theta$ directions into account, these source terms generate the radial and the rotational velocities in the jet. In comparison to Fig. \ref{fig:jetlofac}, one can see that the jet shows more structures along the head of the jet. In the initial stages of the evolution, the jet is denser than the ambient medium. The density profile of the jet gradually decreases in the region up to approximately $z=50$, and the jet is noticeably denser than the surrounding medium. The density contours in Fig. 
 \ref{fig:allmoms_dens} show that as the jet expands beyond a distance of $200\rg$, the jet becomes lighter than the ambient medium. 
Figure \ref{fig:allmoms_dens}-(a) shows that as the jet propagates in the ambient medium, it starts to expand in the lateral direction, but this lateral expansion is halted after some time, which is visible from \ref{fig:allmoms_dens}-(b), this is due to the pressure gradient between the ambient medium and jet beam \citep{pe95}. In Fig. \ref{fig:pres}, we have plotted the contours of pressure for jet Model-B. Initially, we injected an overpressured jet beam into the surrounding medium, but the jet expands very quickly, and the pressure inside the beam decreases. The jet beam creates an overpressured region in front, as seen from panels \ref{fig:pres}(a) and (b). The pressure contours indicate that the medium surrounding the jet beam has a higher pressure than the inside of the beam. Eventually, this high-pressure region quenches the jet beam and halts the lateral expansion. The pressure and density in the jet beam near the compressed region increase as a consequence, compressed and rarefied regions adjacent to each other can be seen in the jet throughout its evolution. The morphology of the jet in the region above $z=100$ resembles that of a supersonic jet moving through a denser medium, as observed in special relativistic simulations. This is expected as the jet has become supersonic, resulting in the formation of shocks, rarefaction fans, and other features commonly observed in supersonic flow. \\   

\begin{figure*}
\begin{center}
	\includegraphics[width=16cm,height=10cm]{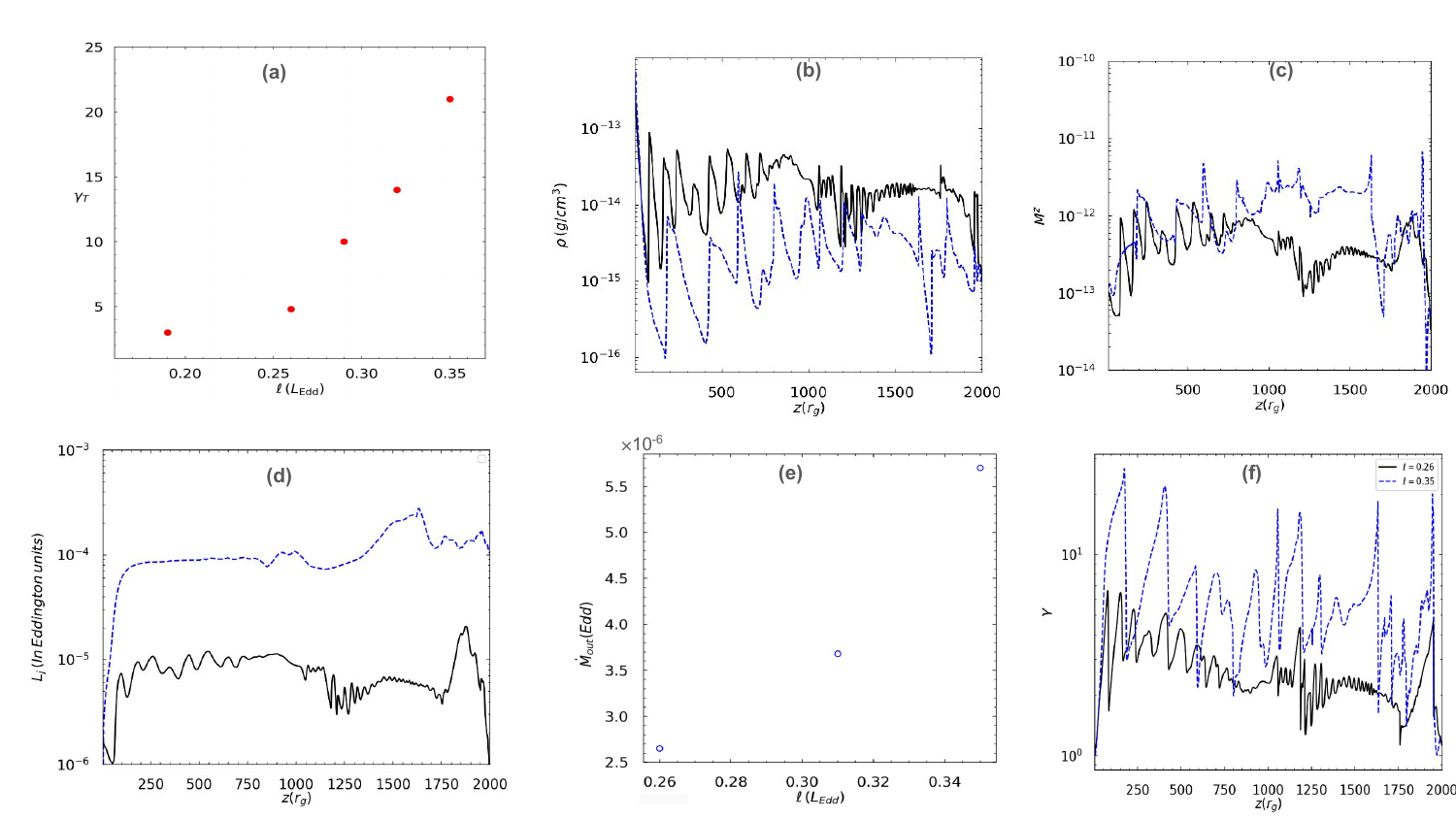}
\end{center}    
    \caption{Panel-(a): Lorentz factor of the jet head as a function of the disk luminosity. Variation mass density (panel-(b)), jet momentum flux along the z direction (panel-(c)), jet kinetic luminosity evaluated along the direction of propagation (panel-(d)), mass outflow rates (panel-(e)), and jet Lorentz factor (panel-(f)) are plotted for two models with $\ell=0.26$ and $\ell=0.35$. Densities are estimated assuming a $10M_{\odot}$ black hole.}
    \label{fig:lofac}
\end{figure*}

In Fig. \ref{fig:lofac} a, we have plotted the Lorentz factor of the jet head at $z\sim 2000\rg$, as a function of the accretion disk luminosity $\ell$. It is evident from the figure that the jets become faster as the disk becomes more luminous. Interestingly, the Lorentz factor is not a linear function of luminosity.
We have also plotted other quantities like the mass density (panel-b), jet momentum flux along the z direction (panel-c), jet kinetic luminosity evaluated along the direction of propagation (panel-d), and mass outflow rates (panel-e) and the jet Lorentz factor along the jet spine (panel-f) for jet models $\ell=0.26$ (solid, black) and $\ell=0.35$ (dashed, blue). The different line styles and colors represent the jet models, driven by the different disk luminosities mentioned in the panel-(f) legend. The densities, mass outflow rate, and jet kinetic luminosities are estimated assuming a $10M_{\odot}$ black hole. All the jet models have the same injection parameters as mentioned in section \ref{sec:setup}. We have calculated the jet kinetic luminosity $L_j$ and mass outflow rate as
\begin{equation}
L_j=\int_{r=0}^{r_j} \gamma^2\rho h v_z 2\pi r dr    
\end{equation}
\begin{equation}
{\dot M}_{\rm out}=\int_0^{r_j} \rho(r,z) u^z(r,z) 2\pi rdr
\end{equation} 
In the above-mentioned equations, the integration is in the radial coordinate and is evaluated at various $z$ for jet kinetic luminosity, and the integration limit is from the $z$ axis (i.e., $r=0$ to the radius of the jet $r_j$).\\
The first thing to notice is that jets that are more accelerated are, on average, less dense. In fact, the jet, driven by disk radiation corresponding to $\ell=0.35$ Eddington
luminosity is orders of magnitude rarer in density. Also, we plot the momentum density (panel-c), which is higher for higher-speed jets. The jet kinetic luminosity is plotted in panel-(d). The disk luminosities for these models are mentioned in the legends (in Eddington units); it is clear from the plot that the jet kinetic luminosity is at least three orders of magnitude lower than the disk luminosities considered. The variation of jet kinetic luminosity is calculated at different locations along the direction of jet propagation to show that it is never higher than the luminosity of the accretion disk for all of the models. Additionally, we must also point out that the ratio of jet kinetic luminosities is also dictated by the distribution of mass densities in addition to the terminal Lorentz factors achieved. The mass outflow rate calculated at $z=2000\,r_g$ (which is the outer edge of the domain) is plotted in panel (e) for different jet models. The mass outflow rates are the order of $\sim 10^{-6}$ Eddington rate. The outflow rates are low because the densities at the jet base are low.

In this paper, we have considered a pair-dominated jet with a very tiny fraction of protons. Since radiative acceleration depends on the total electron (and positron, if present), so it helps in achieving impressive acceleration. The main reason is that if the composition of the jet is mostly pairs, then the opacity is dominated by $\sigma_T/m_e$, but if it is electron-proton, then the opacity becomes $\sim \sigma_T/(m_p+m_e)$. Moreover, the number of leptons also decreases if it is an electron-proton jet. Therefore, the electron-proton jet will not be accelerated to relativistic speeds. In Appendix \ref{app:baryonjet}, we plot the evolution of an electron-proton jet in order to underline the point. The electron-proton jet although is transonic, but only achieves speeds up to $0.2-0.4c$ (Fig. \ref{fig:xi1jet} \& \ref{fig:xi1jet_vz}), and the jet profile is also smooth and has almost no shock or rarefaction fan!

\subsection{Effect of Radiation pressure on collimation}
\begin{figure*}
	\includegraphics[width=14cm,height=16cm]{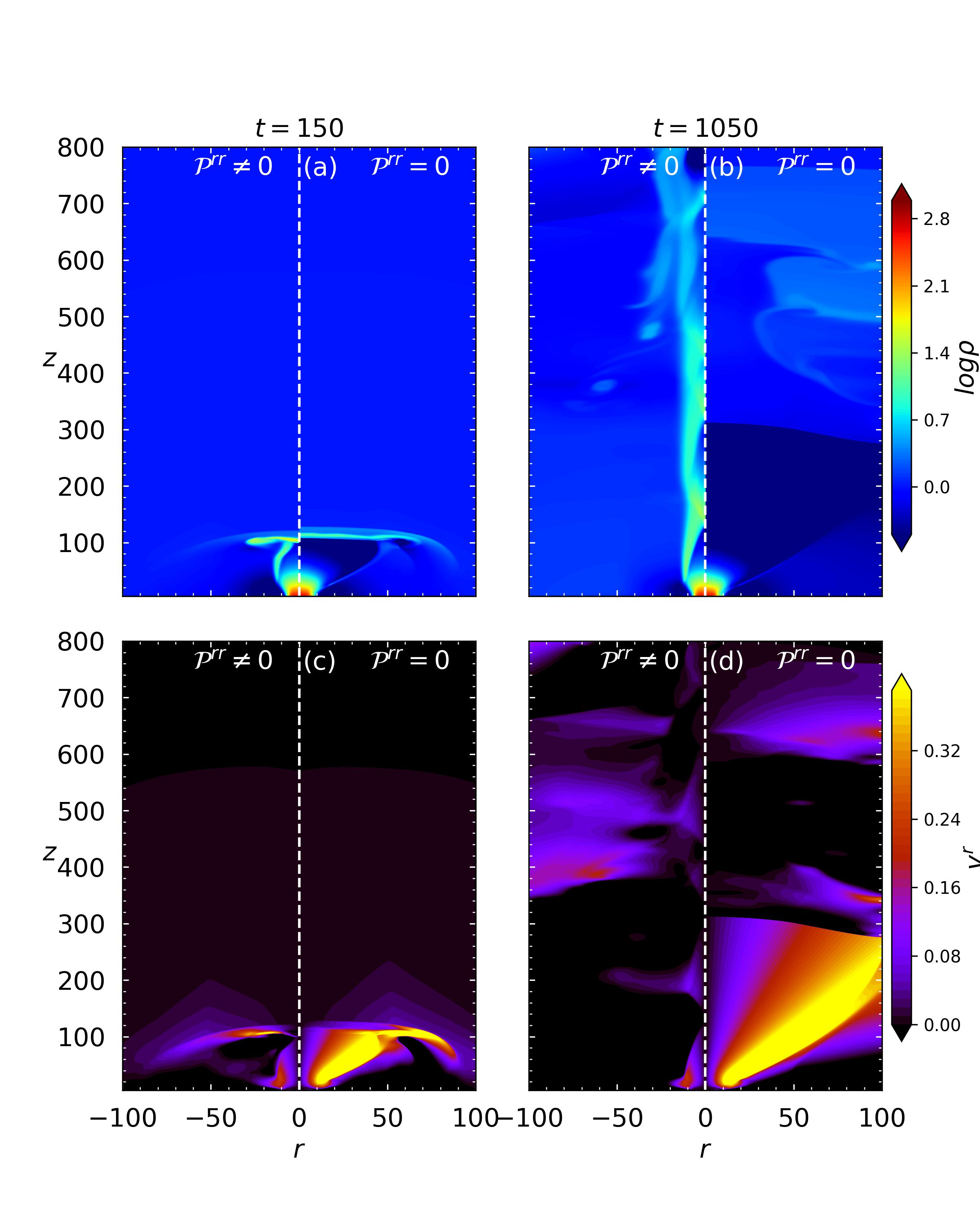}
    \caption{Comparison between the density contours and contours of $v^r$ for jet models with and without $\mathcal{P}^{rr}$.}
    \label{fig:collim1}
\end{figure*}

The collimation of a jet is represented by the value of its radial three-velocity component $v^r$, where a higher value of $v^r$ represents a poor collimation. The generation of $v^r$ is influenced by the radiation flux along the $r$ direction or $\mathcal{F}^r$, but the $\mathcal{P}^{rr}$ component of radiation pressure will try to suppress it. To assess the impact of $\mathcal{P}^{rr}$ on jet collimation, two sets of simulations were compared, and the results are shown in Fig. \ref{fig:collim1}.
Figure \ref{fig:collim1} a \& b displays the density contours
at two different time steps for these simulations. The corresponding contours of $v^r$ are compared in Figs. \ref{fig:collim1} c \& d. In each panel (a)-(d) of Fig. \ref{fig:collim1}, the left half of the panel represents the evolution of the jet when $\mathcal{P}^{rr}$ is considered, while the right half of the panel shows the simulation when $\mathcal{P}^{rr}$ is forcibly put equal to zero. The simulations clearly show that the jet is more collimated under the influence of $\mathcal{P}^{rr}$. Moreover, the collimated jet has also traveled a larger distance along the $z$ direction, as it is faster in comparison to the case when radiation pressure along the $r$ direction is not considered.  The simulations where $\mathcal{P}^{rr}$ is not considered show a higher value of $v^r$ from the very early epochs. As the jet material evolves in time, $v^r$ is limited to a lower value where the drag is considered in the simulation. However, ignoring $\mathcal{P}^{rr}$, results in a continuous increment in $v^r$. Which leads to the jet spreading along the $r$ direction.
Overall, these results suggest that the collimation of a jet is influenced by the interplay between its radial velocity component, $v^r$, and the $\mathcal{P}^{rr}$ component of radiation pressure. Since $v^r$ generated by $\mathcal{F}^r$ is less than the amount of $v^r$ reduced by $\mathcal{P}^{rr}$, one can say that the radiation field above an advective accretion disk can accelerate as well as collimate the jet. 
We must highlight the fact that in Fig. \ref{fig:jetlofac}, the jet also remained collimated without the inclusion of $P^{rr}$. However, The jet collimation mechanisms in Fig. \ref{fig:jetlofac} and Fig. \ref{fig:collim1} are essentially for two different scenarios. In Fig. \ref{fig:jetlofac}, we study the jets with only $E$, $F^z$, and $P^{zz}$ components of the radiation tensor (all measured along the axis of symmetry) and ignoring all other 7 components of radiation tensor, which mimics the dynamics of one-dimensional analytical result in 2D simulations. In Fig. \ref{fig:jetlofac}, as we do not include any force terms along $r$, direction, the jet is only expected to expand along the $r$, only by the virtue of thermal pressure. This expansion will take place through the eigenmodes, which propagate at the local sound speed. However, the acceleration in $z$ direction is important, and the jet becomes supersonic with a few Schwarzschild radii above the jet base. A fast transonic jet does not allow thermal disturbance to have significant time for lateral expansion, and the jet is collimated. 
On the other hand, in Fig. \ref{fig:collim1}, $F^r$ is included as a source term; it acts as an additional source term, which increases $v^r$. Hence, the jet keeps on gaining lateral speed and is poorly collimated. The major role of $P^{rr}$ is to provide the drag along $r$ direction, which may collimate the jet.

\subsection{Rotating jets}

\begin{figure*}
\begin{center}
	\includegraphics[width=9cm,height=12cm]{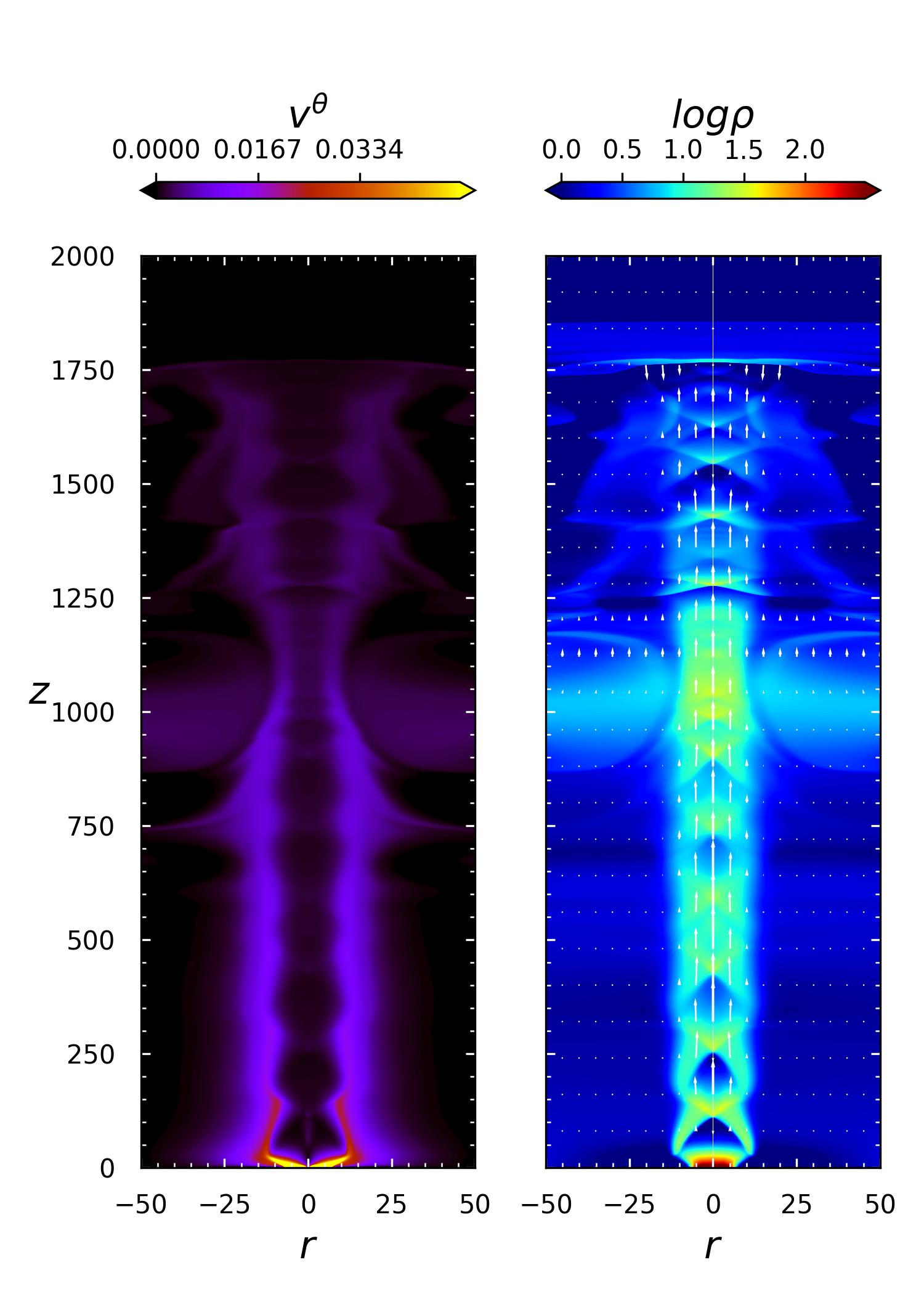}
\end{center}    
    \caption{Contours of $v^\theta$ and $log \rho$ for a rotating jet model plotted at $t=2400\,t_g$.}
    \label{fig:rot}
\end{figure*}

As the jets in AGN and X-ray binaries originate from the accretion disk, they can also contain some rotational velocity to begin with. If the jet has some rotational velocity, it will generate a centrifugal force, which tends to push the matter away from the jet axis and may destroy the collimation. This is evident from Eq. \ref{eq:momentum_r}, as the second term on the RHS generates $v^r$ if $v^\theta$ is non-zero. Hence, the rotating jets may spread along the $r$ direction. As discussed earlier, the distribution of the radiative moments due to PSD is such that the radiation drag along the $\theta$ direction is high in comparison to the source terms so that $v^\theta$ is reduced to a very low value. In this section, we investigate the case of whether a non-zero initial value of $v^\theta$ destroys the collimation or not. In addition to injection velocity along the $z$ direction, we also supply $v^\theta$ to the jet beam. The injection value is taken to be $v^\theta=0.1$, and the rest of the parameters are kept the same. In Fig. \ref{fig:rot}, we have plotted the contours of $v^\theta$ at $t=2400\, t_g$ to show the distribution of the rotational velocity of the jet. It is evident that $v^\theta$ is reduced to a very small value through interaction with the radiation field. Hence, even if the jet originates from the disk with some angular momentum, the angular momentum of the jet beam will be dissipated due to the radiation drag. The value of $v^\theta$ decreases within a few $r_g$, which means the radiation field of the inner PSD itself is the primary agent that takes away the angular momentum of the jet beam. We have also plotted the density contours along with the vector field of momentum flux, which clearly indicates that the jet morphology is not affected significantly as the angular momentum of the jet is dissipated within a very short distance. 

\section{Discussion and Conclusions}
\label{sec:concl}
This paper focuses on investigating the dynamics and morphology of relativistic jets that are propelled by the radiation field of the accretion disk. The jet is accelerated by the momentum imparted by the radiation field, and to understand the jet dynamics, we employ relativistic radiation hydrodynamic equations of motion and a relativistic equation of state. Most of the jet simulations in the relativistic hydrodynamic regime entail investigations of jet morphology with supersonic injections. In this work, we injected subsonic jets with non-relativistic injection speeds. The subsonic jet is also denser than the ambient medium at the base. This setup mimics a jet near its base and is in contrast to
the setup of most relativistic jet simulations. The jet propagates outward against the gravity of the central black hole
but is powered by the momentum deposition of radiation from the accretion disk. It may be noted that gravity is introduced through the $g_{tt}$ term of the metric, somewhat like a weak field approximation in general relativity. In this manner, the tensor property of the equations of motion is maintained, and thus we can
avoid the complication of computing radiative moments by
considering the geodesics of the photons. Since jets travel large distances away from the central object, except
the jet base, which is a few $\rg$ from the central black hole, the geometry is essentially flat in most of the jet. So, treating gravity as a weak field approximation compromises very little in terms of physics, and it is
mathematically consistent.

The results of our simulations clearly show that a jet launched with a subsonic inner boundary condition can attain relativistic terminal velocities. Additionally, the solutions for the jets exhibit transonic characteristics. First, we compare two-dimensional (2-D) jets with one-dimensional (1-D) ones. For 1-D jet (along the $z$ axis), only three radiative moments ($\mathcal{E},~\mathcal{F}^z~\&~\mathcal{P}^{zz}$) are important. In the comparison study, we also used only these three moments, even in 2-D jet simulation.  
In the 1-D case, we produced transonic, relativistic jets. The 2-D jet simulation also produced transonic flow.
We compared the velocity profiles of the two cases along the
axis of symmetry. The two profiles matched up to the first few tens of Schwarzschild radii, where the 2-D jet propagates in a highly collimated geometry. However, as it expands laterally at larger distances and creates additional structures, the terminal velocity achieved is less than the 1-D estimates.
This comparison with the 1-D and the 2-D jet is intended to be mostly
instructive. We wanted to show that both the 1-D and 2D jets produce relativistic and transonic jets. In order to do that, the 2-D jet is not acted on by all the radiative moments.  
The mismatch between 1-D and 2-D jet terminal Lorentz factors had nothing to do with radiation drag and is caused by the interaction of the 2-D jet with the ambient medium 
 and resulting backflow. This mismatch is not due to the inefficiency of radiative interaction to impart momentum. 

In order to ascertain how important radiative driving is, we compared a purely thermally driven jet and a jet driven by radiation, where both are launched with the same subsonic injection parameters. The acceleration of a thermally driven jet is due to the conversion of thermal energy to kinetic energy.
Since the jet studied in this paper is launched with subsonic conditions, the lateral expansion is significant, leading to instabilities due to the growth of the mixing layer.
On the contrary, the deposition of radiative momentum by all the components of radiative moments has a great impact on the acceleration of the jet. In addition to the acceleration, we must also highlight the fact that radiation also helps in making the jets stable against the instabilities due to the growth of the mixing layer. It may be noted that in this comparison, we used all the radiative moment components (ten independent ones), but the magnitude is reduced to ten percent of the actual. And yet, the jet is remarkably stabilized into a collimated flow. 

Simulations of jets with all the radiative moments show a more complex morphology in comparison. The forward shock and jet head start to form, and even the internal shocks form. Not only the jets are relativistic, but they are remarkably collimated. It might be intriguing that while the radiation fields from many accretion disk models were such that the radiation
drag produced by them imposed a mildly relativistic upper limit on jet speed, while the disk model we chose produces a radiation field that accelerates jets to relativistic terminal speeds, and that too for sub-Eddington luminosities. 
Since in this paper, we have used a fully relativistic treatment of the interaction of radiation with optically thin jets, the effect of the radiation drag is automatically included through the mathematical structure. Most of the disk models produce a radiation field that imposes a strict
upper limit for the jet speed that can be achieved by the action of radiation.
It may be noted that the radiation drag effect is essentially the effect of an extended source. The radiative energy density and components of flux and pressure are defined in equations \ref{eq:eng_rad}, \ref{eq:forc_rad} \& \ref{eq:pres_rad}, which are various moments of the intensity. Close to the accretion disk, the directional cosines will play an important role, and the values of these moments at small $z$ will differ from each other. The radiative energy density and radiation pressure terms combine with velocity components and act as drag terms (Eqs. \ref{eq:gr}---\ref{eq:Fico}), while the radiative flux acts as an accelerating agent. Therefore, in regions closer to the disk, if the speed of the jet goes beyond a certain speed limit, then radiative deceleration sets in. The exact value of the 
upper limit is determined by the comparative strength of $\mathcal{E}$, $\mathcal{P}^{ij}$ and $\mathcal{F}^i$. As the jet leaves the accretion disk to a larger distance, the source of the radiation, i.e., the accretion disk, becomes smaller and approaches a point-like source. This 
is shown in Fig. \ref{fig:psd_moms}(k), where $\mathcal{E}\approx ~\mathcal{F}^i \approx \mathcal{P}^{ij}$ which is a hallmark of radiation field of a point source since the directional cosines make less and less contribution. When that happens, the upper limit of speed that could be achieved by the plasma driven by the radiation is $c$ --- the speed of light. Our disk model has two parts: an inner torus-like region
and a slimmer outer sub-Keplerian region. Both these components are sources of radiation, and the inner PSD is brighter than the SKD. However, radiation fields from these two components maximize at a short distance
above the disk surface. However, at distances $\gsim 100 \rg$, the disk starts to behave like a point source, but the magnitude of the combined radiation field
remains significant due to the addition of the radiation field from SKD. Therefore, the radiation field from this disk
model can accelerate the jet to relativistic speeds.
We plot the terminal Lorentz factor $\gamma_{\small {\rm T}}$ with the disk luminosity and show that even multidimensional jet simulations can produce relativistic jets. Besides the jet acceleration, our simulations also show
that radiation collimates and stabilizes the jet. It is common knowledge that radiation naturally spreads. Therefore, jet collimation by disk radiation appears counterintuitive. One of the major reasons that the jet cannot spread laterally is because of the impressive acceleration in the $z$ direction. The drag terms are weakest along the direction of propagation of the jet; hence, the jet is accelerated along this direction. On the other hand, the drag along the $r$ and $\theta$ direction is stronger and limits $v^r$ and $v^\theta$ to smaller values, which results in a jet beam propagating vertically up with a very high velocity and remains fairly collimated. Hence, the radiation also acts as a collimating agent. In particular, the $P^{rr}$ component of the radiation pressure is the principal collimating agent, which reduces the velocity along the $r$ direction. In order to check the effect of $P^{rr}$ on collimation, we compared two sets of simulations. We found out that in the absence of $P^{rr}$, the value of $v^r$ keeps on increasing, the jet expands along the lateral direction, and the collimation is very poor. We also investigated the jets that start with some initial rotational velocity. As these jets propagate through the radiation field of the disk, they lose the angular momentum as the radiation drag terms are much stronger in comparison to the source terms that can generate the rotational velocity component in the jet. Hence, even if jets start with some initial angular momentum, it will be dissipated through the interaction with the radiation field. The jet will be stable against the centrifugal force that can be generated through the rotational velocity component. One can, therefore, conclude that an astrophysical jet, which is being driven by the radiation field of an advective type of accretion disk, can be collimated and accelerated starting from subsonic, non-relativistic speeds at the jet base. The terminal Lorentz factors can range from a few to a few tens if the accretion disk luminosity is about ten percent to a few tens percent of the Eddington luminosity.

It would be interesting to study the radiative acceleration model of the jet in tidal disruption events and also in highly energetic gamma-ray bursts-like scenarios as these events generate an intense radiation field. We have identified these aspects as potential future prospects and will be reported elsewhere. \\

\begin{acknowledgments}
A.T. was supported in part by National Science Foundation (NSF) PHY-2308242 and OAC-2310548, to the University of Illinois at Urbana-Champaign and acknowledges support from the National Center for Supercomputing Applications (NCSA) at the University of Illinois at Urbana-Champaign through the NCSA Fellows program. The authors also acknowledge the anonymous referee for fruitful suggestions.
\end{acknowledgments}

\appendix
\section{Radiative Moments}
The radiative moments are calculated as
\label{app:radmom}

\begin{subequations}
\begin{equation}
E_{\rm rd}=\frac{1}{c}\left(\int I_{\rm ps} d\Omega_{\rm ps} +\int I_{\rm sk} d\Omega_{\rm sk}\right)     
\label{eq:eng_rad}
\end{equation}    

\begin{equation}
F_{\rm rd}^i=\int I_{\rm ps} l^i d\Omega_{\rm ps} +\int I_{\rm sk} l^i d\Omega_{\rm sk}     
\label{eq:forc_rad}
\end{equation}    

\begin{equation}
P_{\rm rd}^{ij}=\frac{1}{c}\left(\int I_{\rm ps} l^il^j d\Omega_{\rm ps} +\int I_{\rm sk} l^il^j d\Omega_{\rm sk}\right)      
\label{eq:pres_rad} \end{equation}
\end{subequations}

In equations (\ref{eq:eng_rad})-(\ref{eq:pres_rad}), $I_{\rm ps/\rm sk}$ represents the frequency integrated intensity from the disk components (PSD/SKD), $d\Omega$ is the solid angle, and $l^i$s are the direction cosines. For SKD, we assume that the synchrotron emission is the dominant emission mechanism and the intensity is given as 

\begin{equation}
I_{\rm sk}=\left[\frac{16}{3}\frac{e^2}{c}\left(\frac{eB_{\rm sk}}{\mel c}\right)^2\Theta_{\rm sk}^2n_{\rm sk}\right]\frac{\left(d_0\,{\rm sin\theta_{sk}}+r\,{\rm cos\theta_{sk}}\right)}{3}\,{\rm erg\, cm^{-2} s^{-1}}
\label{eq:skd_synch}    
\end{equation}

where $B_{\rm sk},\,\Theta_{\rm sk},\,n_{\rm sk}$ represent the magnetic field, local dimensionless temperature, and electron number density, respectively. We assume a stochastic magnetic field in SKD with a constant magnetic to gas pressure ratio $p_{\rm mag}=\beta{p_{\rm gas}}$. $\theta_{sk}$ is the semi-vertical angle for the SKD, and $d_0$ is the intercept of the SKD on the jet axis taken as $d_0=0.4\hs$. \\
$\hs$ represents the height of PSD at its outer edge ($\xs$). The height of PSD at $\xs$ is taken as $\hs=0.6(\xs-1)$.\\

For PSD, the intensity is calculated as $I_{\rm ps}=l_{\rm ps}L_{\rm Edd}/\pi A_{\rm ps}$, where $A_{\rm ps}$ is the surface area of the PSD and $l_{\rm ps}$ is the luminosity of the PSD in units of the Eddington luminosity ($L_{\rm Edd}$). \\

The location of the outer edge of PSD is calculated from the SKD accretion rate $\dot{m}_{\rm sk}$ \cite{vkmc15}

\begin{equation}
\xs=64.8735-14.1476 \dot{m}_{sk}+1.242\dot{m}_{sk}^2-0.0394\dot{m}_{sk}^3
\label{eq:xs_mdot}
\end{equation}

\section{Electron-proton jet in Radiation field}
\label{app:baryonjet}

\begin{figure*}
\begin{center}
	\includegraphics[width=18cm,height=7cm]{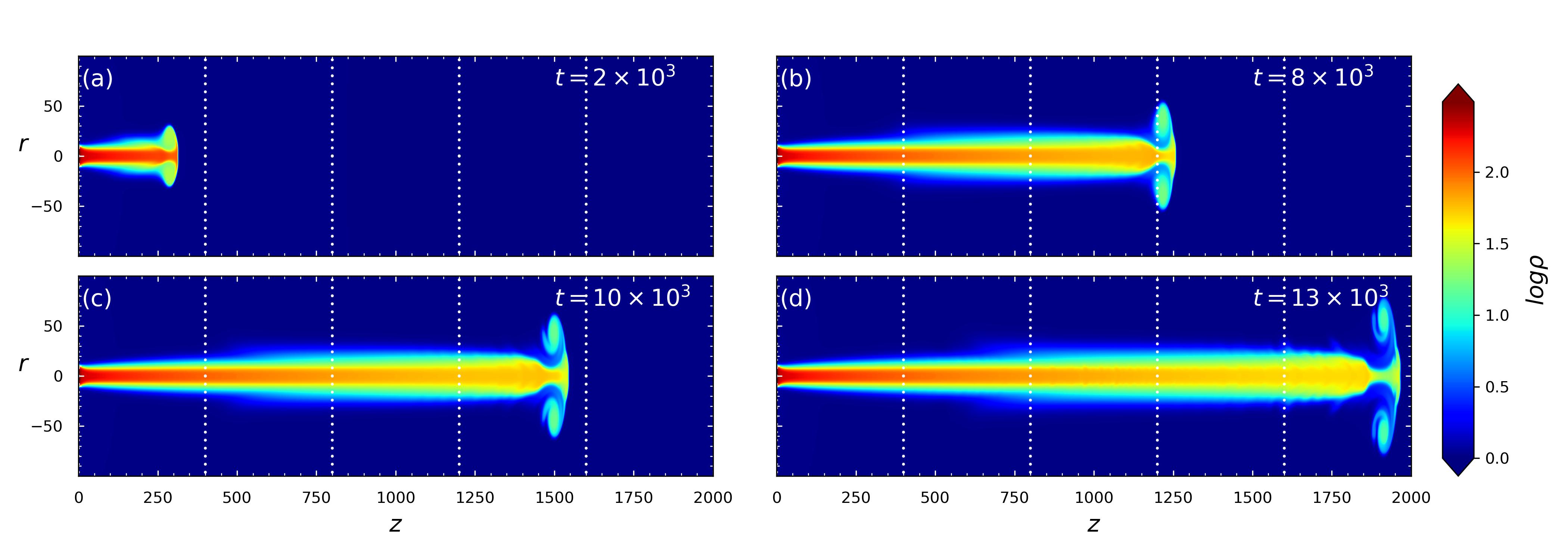}
\end{center}    
    \caption{Density contours for an electron-proton jet at various time steps. The injection parameters and radiation field configuration are similar to the one considered in Fig. \ref{fig:allmoms_dens}.}
    \label{fig:xi1jet}
\end{figure*}

The amount of momentum transferred from the radiation field to the jet directly depends on the lepton density of plasma, as shown in Eq. \ref{eq:gico}. Hence, for the same disk parameters and injection parameters for the jet beam, the electron-proton ($\xi=1.0$) jet will be slower in comparison to lepton-dominated jets. In Fig. \ref{fig:xi1jet}, we have plotted the density contours for an electron-proton jet at different time steps. If we compare the time stamps mentioned in the panels Fig. \ref{fig:xi1jet}(a)-(d) with  Fig. \ref{fig:allmoms_dens}(a)-(d), we can easily conclude that the jet with plasma composition $\xi=1.0$ is very slow in comparison to lepton dominated jet. Apart from the difference in speed, there are remarkable differences in overall morphological features. The electron-proton jet is slower and heavier and shows relatively lower turbulent features as it propagates through the ambient medium. The vortices are confined near the jet head region only. The internal structure of the beam also shows a lot of difference; while the lepton-rich jet beam shows the presence of multiple internal shocks and rarefactions, these features are absent in the electron-proton jet beam, which is also evident from the contours of $v^z$ as the velocity profile shows a smooth variation along the direction of propagation. The maximum velocity in the jet beam is always less than $0.3c$ for the electron-proton jet. The vortices alongside the jet beam that are prominently seen in the case of the electron-positron jet (Fig. \ref{fig:allmoms_dens}) do not develop in the case of the electron-proton jet.     

\begin{figure*}
\begin{center}
	\includegraphics[width=18cm,height=7cm]{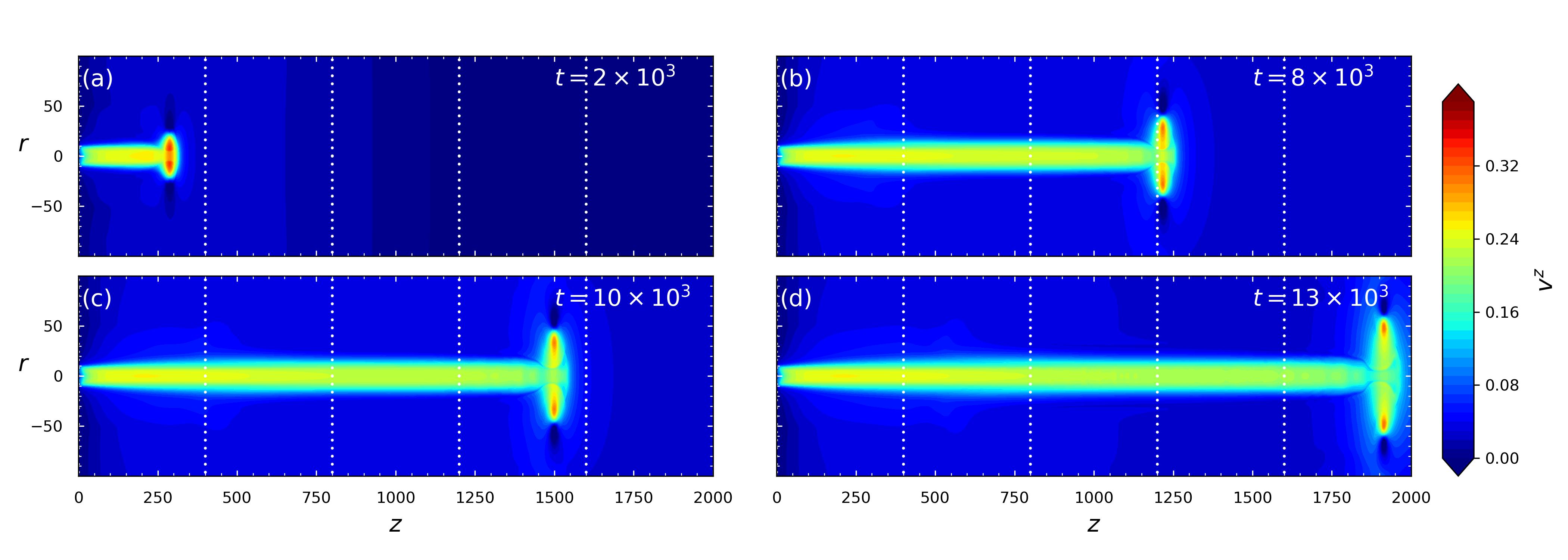}
\end{center}    
    \caption{Contours of $v^z$ for an electron-proton jet.}
    \label{fig:xi1jet_vz}
\end{figure*}
\bibliography{biblio}{}
\bibliographystyle{aasjournal}

\end{document}